\documentclass[twocolumn,showpacs,preprintnumbers,amsmath,amssymb,superscriptaddress]{revtex4}

\usepackage{graphicx}
\usepackage{graphics}
\usepackage{dcolumn}
\usepackage{bm}

\begin{document}

\title{Thermal Expansion of the Heavy-fermion Superconductor PuCoGa$_5$}

\author{R. Eloirdi}
\affiliation{European Commission, Joint Research Centre (JRC), Directorate for Nuclear Safety and Security, Postfach 2340, D-76125 Karlsruhe, Germany}

\author{C. Giacobbe}
\affiliation{European Synchrotron Radiation Facility (ESRF), B.P.220, F-38043 Grenoble, France}

\author{P. Amador Celdran}
\affiliation{European Commission, Joint Research Centre (JRC), Directorate for Nuclear Safety and Security, Postfach 2340, D-76125 Karlsruhe, Germany}

\author{N. Magnani}
\affiliation{European Commission, Joint Research Centre (JRC), Directorate for Nuclear Safety and Security, Postfach 2340, D-76125 Karlsruhe, Germany}

\author{G. H. Lander}
\affiliation{European Commission, Joint Research Centre (JRC), Directorate for Nuclear Safety and Security, Postfach 2340, D-76125 Karlsruhe, Germany}

\author{J.-C. Griveau}
\affiliation{European Commission, Joint Research Centre (JRC), Directorate for Nuclear Safety and Security, Postfach 2340, D-76125 Karlsruhe, Germany}

\author{E. Colineau}
\affiliation{European Commission, Joint Research Centre (JRC), Directorate for Nuclear Safety and Security, Postfach 2340, D-76125 Karlsruhe, Germany}

\author{K. Miyake}
\affiliation{Toyota Physical and Chemical Research Institute, Nagakute, Aichi 480-1192, Japan}

\author{R. Caciuffo}
\affiliation{European Commission, Joint Research Centre (JRC), Directorate for Nuclear Safety and Security, Postfach 2340, D-76125 Karlsruhe, Germany}

\date{\today}

\begin{abstract}
We have performed high-resolution powder x-ray diffraction measurements on a sample of $^{242}$PuCoGa$_{5}$, the heavy-fermion superconductor with the highest critical temperature $T_{c}$ = 18.7 K. The results show that the tetragonal symmetry of its crystallographic lattice is preserved down to 2 K. Marginal evidence is obtained for an anomalous behaviour below $T_{c}$ of the $a$ and $c$ lattice parameters. The observed thermal expansion is isotropic down to 150 K, and becomes anisotropic for lower temperatures. This gives a $c/a$ ratio that decreases with increasing temperature to become almost constant above $\sim$150 K. The volume thermal expansion coefficient $\alpha_{V}$ has a jump at $T_{c}$, a factor $\sim$20 larger than the change predicted by the Ehrenfest relation for a second order phase transition. The volume expansion deviates from the curve expected for the conventional
anharmonic behaviour described by a simple Gr\"{u}neisen-Einstein model. The observed differences are about ten times larger than the statistical error bars but are too small to be taken as an indication for the proximity of the system to a valence instability that is avoided by the superconducting state.
\end{abstract}

\maketitle

\section{Introduction}
PuCoGa$_{5}$ has the highest $T_{c}$ (18.7 K) of any heavy-fermion superconductor. Fifteen years after its discovery \cite{sarrao02} our understanding of much of this material remains at best confused \cite{curro05,jutier08,flint08,hiess08,daghero12,bauer12,ramshaw15}. We do know from NMR \cite{curro05} and point-contact spectroscopy \cite{daghero12} measurements that the superconducting state has $d$-wave symmetry. Magnetic form-factor measurements with polarized neutron diffraction \cite{hiess08} have shown that the ground state is not the conventional $5f^{5}$ state found in many Pu intermetallics. Neutron inelastic scattering has failed to detect any sign of a resonance as found, for example, in the isostructural CeCoIn$_{5}$ compound \cite{stock12}, which also has $d$-wave symmetry, although the difficulty of performing these neutron experiments on Pu should not be overlooked. Recent theoretical efforts \cite{flint11,pezzoli11,zhu12} have concluded that the driving mechanism for superconductivity is valence fluctuations. Electronic structure calculations combining the local-density approximation with an exact diagonalization of the Anderson impurity model \cite{shick13} show an intermediate $5f^{5}-5f^{6}$-valence ground state and delocalization of the $5f^{5}$ multiplet of the Pu atom $5f $ shell. The $5f $ local magnetic moment is compensated by a moment formed in the surrounding cloud of conduction electrons, leading to a singlet Anderson impurity ground state.

The presence of valence fluctuations has been recently suggested by resonant ultrasound spectroscopy measurements, showing that the three compressional elastic moduli exhibit anomalous softening upon cooling, which is truncated at the superconducting transition \cite{ramshaw15}. These results have been interpreted as evidence for a valence transition at a $T_{V} < T_{c}$ that is avoided by the superconducting state, suggesting that PuCoGa$_{5}$ is near a critical-end point involved in the unconventional superconductivity \cite{ramshaw15}. However, the identification of the fluctuating order parameter responsible for the observed anomalous softening requires information on the thermal expansion of the lattice, which is not available. Moreover, crystallographic studies at low temperature (T) have not yet been performed, so that the occurrence of a lattice distortion at $T_{c}$ (or above $T_{c}$) has not been verified. These are the issues that we have addressed by performing high-resolution x-ray diffraction measurements in the T range between 2 and 300 K on a defect-free polycrystalline sample of  $^{242}$PuCoGa$_{5}$. The results of our investigation show that no measurable structural distortion is associated with the stabilization of the superconducting phase. The thermal expansion is isotropic down to 150 K, and anisotropic for lower temperatures, which is not surprising for a superconductor with an order parameter of $d$-wave  symmetry. The T dependence of both $a$ and $c$ lattice parameters shows small anomalies at $T_{c}$ and a behavior that deviates from the one expected by the simple quasi-harmonic approximation commonly used to describe the thermal expansion in solids. However, no convincing evidence is found for an incipient valence transition of the Pu electronic configuration associated with the formation and condensation of Cooper pairs.

\section{Experimental details and results}
\begin{figure}
\centerline{ \includegraphics[width=7.5cm]{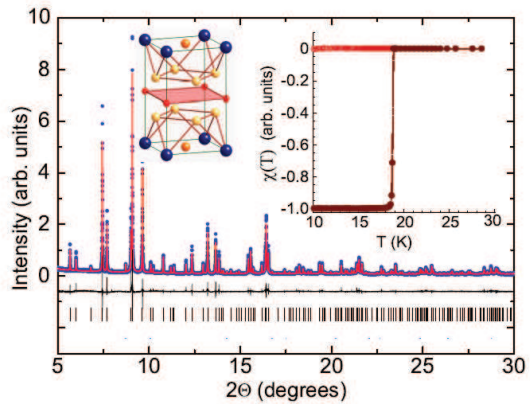}}
\caption{(Color online) X-ray diffraction pattern recorded for
PuCoGa$_{5}$ at 5 K; blue points are experimental data, the red line is the refined Rietveld profile (R$_{wp}$ = 8.77\%, R$_{wp}$/R$_{exp}$ = 1.2). The residual is given by the black line at the bottom; vertical ticks represent the angular position of Bragg peaks. No impurity phases were detected. Left inset: crystallographic unit cell (S. G. P4/$mmm$, No. 123). Pu atoms (blue spheres) occupy the 1$a$ Wyckoff position, Co atoms (red spheres) are located at the 1$b$ position, halfway between the Pu
atoms along the $c$ direction, and the Ga atoms sits on two crystallographic positions, one (orange spheres) in the center of the basal planes (1$c$) the other (4$i$) in the rectangular faces of the unit cell (yellow spheres) with the position  (0, 1/2, $z$). The refined value of the $z$ parameter at 5 K is $z$ = 0.3075(2). Right inset: temperature dependence of the magnetic susceptibility measured under zero-field cooling (closed brown circles) and field cooling conditions (open red circles, applied field of 10$^{-3}$ tesla), providing a T$_{c}$ of 18.7 K. \label{diffra}}
\end{figure}

The experiment was performed at the ID22 beamline of the European Synchrotron Radiation Facility (ESRF) in France. Data have been collected on a sample obtained by crushing a single crystal grown at the Karlsruhe establishment of the Joint Research Centre in a Ga flux using the $^{242}$Pu isotope (99.932 wt\% $^{242}$Pu, 0.035 wt\% $^{241}$Pu, 0.022 wt\% $^{240}$Pu, 0.005 wt\% $^{239}$Pu, 0.004 wt\% $^{238}$Pu, 0.002 wt\% $^{244}$Pu on December 2015) to avoid effects from radiation damage and self-heating. The total sample mass was 4.6~mg, corresponding to a plutonium mass of 1.7~mg and a total activity of $\sim$760~kBq. Magnetic susceptibility and specific heat measurements show superconductivity below T$_{c}$ = 18.7 K (see inset of Fig. \ref{diffra}).
Following a protocol developed for powder diffraction measurements at synchrotron radiation sources on other transuranium isotopes \cite{klimczuk12},  the sample was put inside a hermetic holder providing four levels of containment. For this, we used a kapton capillary (1~mm diameter, $\sim$25~mm in length) half filled with Stycast. The resin was allowed to cure, before a 5~mm mixture of a second resin (Epofix) and the sample was inserted with a pipette. The Epofix was used because of the lower viscosity, allowing easier mixing with the powder sample and insertion into the narrow kapton capillary. The remainder of the capillary was then filled with Stycast and, once fully cured, it was inserted into a drilled-out plexiglass rod, which was sealed with a plexiglass plug, glued with further Stycast and finally enveloped within a 4~mm polyimide tube. Due to the contamination risk generated by the plutonium element, all operations of preparation and encapsulation have been carried out in shielded gloveboxes under inert nitrogen atmosphere following well-established safety procedures.

The channel-cut Si-111 monochromator of ID22 provided an incident beam wavelength of 0.354155~\AA. The sample capillary was mounted on the axis of the diffractometer inside a liquid-helium-cooled cryostat allowing reaching a base temperature of 2~K. To avoid any risks of mechanical failure of the containment, the sample was not spun within the cryostat. This did not result in preferred sample orientation issues in the data, as the setting process in the resin eliminates any preferred orientation and provides a good sample average. A NIST 640c Si standard was used to calibrate the Si-111 multi-analyser stage. In the first part of the experiment, the diffraction pattern was measured at several temperatures, from 2 to 300 K, with acquisition times up to four hours. The T dependence of the lattice parameters was obtained from the Rietveld refinement of diffraction patterns collected on warming from 5 K up to 260 K, with a counting time of 1 hour at each temperature. The experimental resolution was of the order of $\Delta d/d$ = 10$^{-6}$. The main results are summarized below.

\begin{figure}
\centerline{ \includegraphics[width=7.5cm]{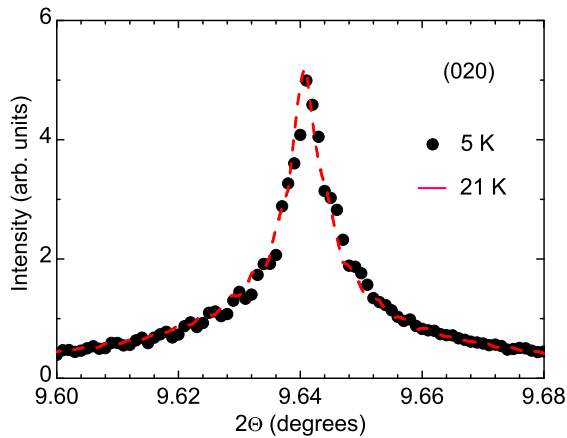}}
\caption{(Color online) The (0 2 0) Bragg peak measured at 5 K (black circles) and 21 K (red dashed line). Neither splitting nor broadening of the lineshape are observable, indicating that the tetragonal symmetry is preserved in the superconducting phase. \label{020line}}
\end{figure}

The best fit of the diffraction profile is obtained within the tetragonal P4/$mmm$ Space Group in the whole temperature range explored in this experiment. Close examination of the shape and width of individual Bragg peaks shows no evidence for the occurrence of a lattice distortion across T$_{c}$, as shown for the (020) Bragg peak in Fig. \ref{020line}.

\begin{figure}
\centerline{ \includegraphics[width=8cm]{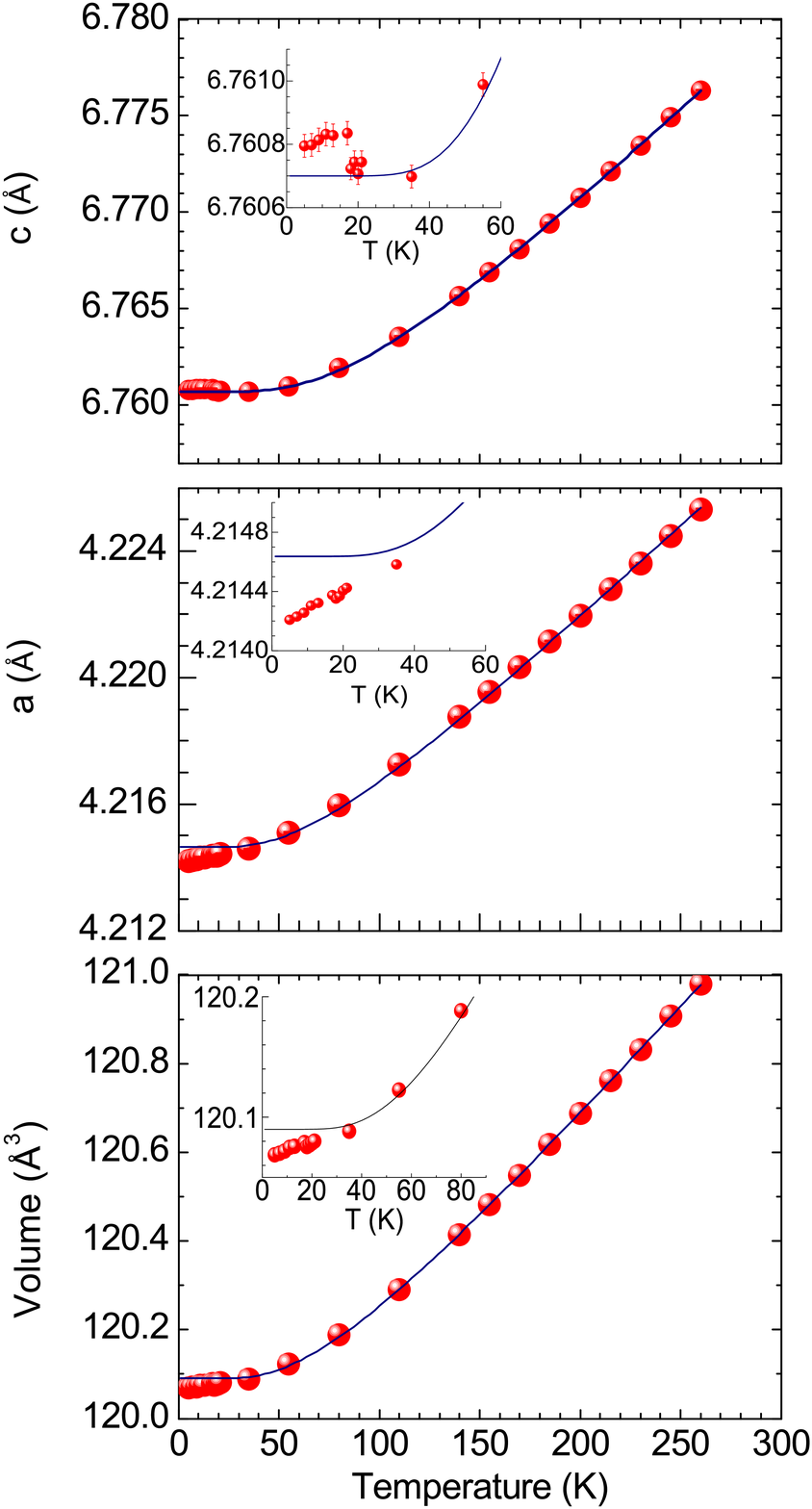}}
\caption{Thermal expansion of the PuCoGa$_{5}$ lattice parameters and unit cell volume. The solid lines represent a fit to a Gr\"{u}neisen-Einstein model, as explained in the text. Vertical bars give the statistical error multiplied by a factor 5. If not visible, error bars are smaller than the symbols size. \label{Tdepac}}
\end{figure}

The temperature dependence of the lattice parameters $a$ and $c$ is shown in Fig. \ref{Tdepac}, together with the thermal expansion of the unit cell volume $V(T)$. The error bar on the experimental data represent the error $\sigma_{R}$ estimated from the Rietveld refinement multiplied by a factor 5. The solid line is a fit to a simple one-phonon Gr\"{u}neisen-Einstein model,

\begin{equation}\label{GE}
  ln\left(\frac{V(T)}{V_{0}}\right) = \frac{k_{B} n}{B V_{m}}\frac{\gamma T_{E}}{exp(T_{E}/T)-1}
\end{equation}

\noindent
and equivalent expressions for the lattice parameters $a(T)$ and $c(T)$. In Eq. \ref{GE}, $V_{0}$ is the unit cell volume at T = 0, $k_{B}$ is the Boltzman constant, $n$ = 7 is the number of atoms per unit cell, $B$ $\simeq$ 89-100 GPa is the bulk modulus \cite{normile05,heathman10,ramshaw15}, $V_{m}$ = 7.23$\times$10$^{-5}$ m$^{3}$/mol  is the molar volume, $\gamma$ is the Gr\"{u}neisen parameter and $T_{E}$ is the Einstein temperature. The best fit is obtained for $V_{0}$ = 120.090(3) {\AA}$^{3}$ (c$_{0}$ = 6.7607(4) {\AA}, a$_{0}$ = 4.2146(3) {\AA}), $\gamma$ = 5.2(4), and $T_{E}$ = 197(4) K. The values estimated by this simple model are in line with those obtained by self-consistent calculations reported in Ref. \onlinecite{filanovich16}. Moreover, the Gr\"{u}neisen parameter has the order of magnitude reported for other mixed-valent Ce and U compounds, for instance CePd$_{3}$, CeSn$_{3}$, and UAl$_{2}$ \cite{thompson94}.

Although the simple model above describes well the behavior at high temperature, clear deviations from the predicted dependence are observed below T$_{c}$. Whilst $a$ decreases linearly with decreasing T, $c$ has a small expansion at T$_{c}$ and becomes constant at lower temperatures, a behavior similar to the one calculated by Millis and Rabe for La$_{2-x}$Sr$_{x}$CuO$_{4}$ and YBa$_{2}$Cu$_{3}$O$_{7}$ by taking into account Gaussian fluctuation corrections to the mean field superconducting free energy \cite{millis88}.
As a consequence, the volume expansion deviates from the curve expected for the conventional anharmonic behavior described by the Gr\"{u}neisen-Einstein model, with differences that are about two times larger than the error bars given by 5$\sigma_{R}$. One must, of course, be aware that Eq. \ref{GE} has his roots in the Einstein approximation for the specific heat of a solid, which underestimates the contribution of long wavelength vibrational modes at very low temperatures. However, as significant differences with the curve predicted by more sophisticated models are expected at temperatures much smaller than the T$_c$ in PuCoGa$_5$, its use to signal an anomaly in the observed experimental data at the onset of superconductivity is, in the present case, justified.

\begin{figure}
\centerline{ \includegraphics[width=7.0cm]{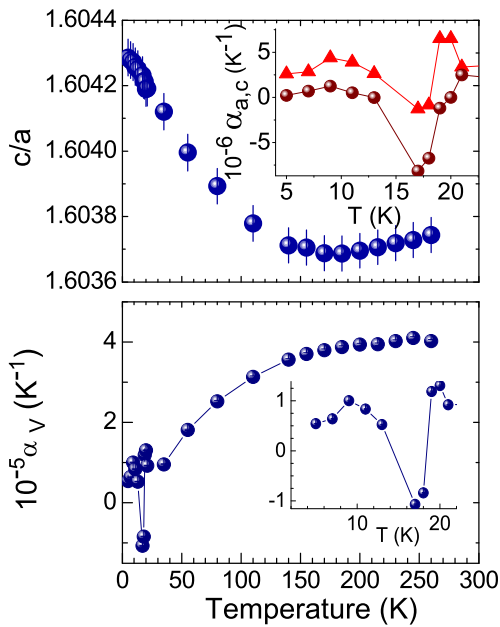}}
\caption{Top panel: temperature dependence of the $c/a$ ratio for PuCoGa$_{5}$. The inset shows the temperature dependence of the linear thermal expansion coefficients along the $a$ (triangles( and $c$ (circles) crystallographic directions. Bottom panel: Thermal expansion coefficient for the unit cell volume (in an expanded scale around T$_c$ in the inset). Error bars are estimated as five times the statistical error provided by the Rietveld refinement and are smaller than the symbol size if not shown. The solid lines are guide to the eye. \label{alphaV}}
\end{figure}

As shown in Fig. \ref{alphaV} (top panel), upon cooling the expansion is isotropic down to 150 K and anisotropic for lower temperatures. This results in a $c/a$ ratio that decreases with increasing T to become almost constant above $\sim$150 K. It is interesting to note that the marked increase of the $c/a$ ratio below 150 K occurs in the temperature range where Ramshaw \textit{et al.} [\onlinecite{ramshaw15}] observe an anomalous softening of the bulk modulus and a significant temperature dependence of the in-plane Poisson ratio.  Such a behavior was attributed in Ref. [\onlinecite{ramshaw15}] to the development of in-plane hybridization between Pu 5$f$ moments and conduction electrons.

The inset (top panel) of Fig. \ref{alphaV} shows the linear thermal expansion coefficients along the $a$ and $c$ directions around T$_c$. The temperature dependence of the volume thermal expansion coefficient $\alpha_{V}$ is shown in the bottom panel of Fig. \ref{alphaV}. The presence of an anomaly with a minimum at T$_{c}$ has been confirmed by repeating the sequence of measurements both on warming and on cooling cycles.

\section{Discussion}
The anisotropic change in thermal expansion at T$_{c}$ is not unexpected for a $d$-wave superconductor adjusting its crystal structure in order to minimize the lattice free energy. On the other hand, the observed deviation of the unit cell volume with respect to the Gr\"{u}neisen-Einstein prediction is much larger than the one obtained from the first Ehrenfest equation for 2$^{nd}$ order phase changes. Such equation relates the difference between the temperature derivative of the volume at constant pressure calculated above and below the phase transition with the jump of the specific heat at T$_{c}$ and the initial slope of the hydrostatic pressure dependence of T$_{c}$ (which measures the average of the stress derivatives)
\begin{equation}\label{Er}
  \left(\frac{\partial V_{s}}{\partial T}\right)_{p} - \left(\frac{\partial V_{n}}{\partial T}\right)_{p} = \Delta \alpha_{V} V_m = \frac{\partial T_{c}}{\partial p} \frac{C_{s}-C_{n}}{T_c}
\end{equation}
\noindent
where $\Delta \alpha_{V}$ = $\alpha_{Vs} - \alpha_{Vn}$ is the difference between the thermal expansion coefficients in the superconducting and normal phase. Previous studies have reported $\partial T_c/\partial p$ = 0.4(2)$\times$10$^{-9}$ K/Pa \cite{griveau14} and $(C_{s}-C_{n})/T_{c}$ = 0.110(4) J mol$^{-1}$ K$^{-2}$ \cite{sarrao02}, leading to $\Delta \alpha_{V}$  = 0.6$\times$10$^{-6}$ K$^{-1}$. This value for the thermal expansion discontinuity is comparable with those calculated for La$_{2-x}$Sr$_{x}$CuO$_{4}$ and YBa$_{2}$Cu$_{3}$O$_{7}$ in Ref.~[\onlinecite{millis88}] but it is smaller by one order of magnitude than the anomaly observed in the experimental curve shown in Fig. \ref{alphaV}. Moreover, the positive jump at T$_c$ of $\alpha_V$ (with increasing T), in conjunction with the negative jump of the specific heat, would indicate an initial negative value for $\partial$T$_c$/$\partial$p, in contrast to the direct measurement.

Such discrepancies in sign or magnitude have been observed in other superconductors like for example the A15 material V$_3$Si \cite{meingast03}, the Chevrel phase PbMo$_6$S$_8$ \cite{meingast03} or the iron-based layered superconductor Ba(Fe$_{1-x}$Co$_{x}$)$_{2}$As$_{2}$ \cite{luz09}, where the thermal expansion is also highly anisotropic and the derivative $\partial$T$_{c}$/$\partial$p deduced from the Ehrenfest relation is negative, whereas the pressure diagram is clearly displaying an increase of T$_c$ with increasing pressure around p $\sim$ 0 GPa. Several hypotheses have been invoked to account for these deviations, but no clear explanation has emerged yet, so we refer interested readers to references therein. One should also notice that PuCoGa$_5$ is a plutonium-based material and Pu element is already at the origin of numerous exotic phenomena, such as a negative thermal dilatation in the $\delta$ phase \cite{shim07}, due to its specific electronic structure  that is far from being fully understood.

Although we do not have any straightforward explanation for the departure from the predictions of the Ehrenfest relation, we think that this is an interesting finding that calls for further studies.  Definitely, determining the thermal expansion of PuCoGa$_5$ single crystals along the main directions with a high-sensitivity technique like dilatometry would be valuable to yield more details on this anomaly at T$_c$ and stimulate theoretical work.

Thermal expansion measurements have been reported for the isostructural heavy-fermion superconductor CeCoIn$_5$ (T$_c$ = 2.3 K) \cite{takeuchi02}. Also in that case, the thermal expansion shrinks for the [100] direction in the superconducting state, while it expands for [001]. The volume thermal expansion decreases linearly down to T$_c$ with decreasing temperature, and more rapidly so in the temperature range from T$_c$ down to 1.5 K \cite{takeuchi02}. This is similar to what we report for PuCoGa$_5$, but for CeCoIn$_5$ the coefficient of the volume thermal expansion shows an anomaly with a lambda shape, which is not the case for PuCoGa$_5$. Moreover, applying the Ehrenfest relation to CeCoIn$_5$ leads to a correct estimate for $\partial$T$_c$/$\partial$p (both in sign and magnitude), again in contrast with what we report for the Pu analogue.

The linear decrease of the unit cell volume with decreasing temperature below T$_{c}$  is qualitatively similar to the one observed in CeRu$_{2}$Si$_{2}$, a compound where the Kondo screening (changing the 4f$^{1}$ localized
state to a non-magnetic 4f-itinerant state) is accompanied by a volume contraction below the Kondo temperature T$_{K}$ = 20 K \cite{hiranaka13,lacerda89}. In that case, the phenomenon can be interpreted in the framework of a theory describing critical valence fluctuations involving 4f$^{1}$ and 4f$^{0}$ electronic configurations \cite{miyake14},
although CeRu$_{2}$Si$_{2}$ is thought to be relatively far from criticality \cite{flouquet05}. According to Ref. [\onlinecite{miyake14}],
the variation of the $f$-shell occupation number $\Delta n_f = n_f(T) - n_f(0)$ is
proportional to the volume change $\Delta V(T)$, $\Delta n_f \propto \zeta \chi_{0} \Delta V / (\eta_{0}+bT^{\xi})$,
where $\zeta$ is a temperature independent constant that relates the energy variation of the f-electrons levels to the volume change,   $\chi_0$ is the non-interacting susceptibility of the order of the quasiparticle density of states, $\eta_0$ is a parameter characterizing the degree of departure from the critical point, and $\xi$ a critical exponent.

Although a theory for valence transitions between electronic configurations with occupation numbers higher than 1 is not yet available, the behavior should be qualitatively similar, and the observed volume shrinkage could be an indication that the valence of the Pu atoms changes below T$_{c}$. However, it must be emphasized that the observed departure of the volume from the Gr\"{u}neisen-Einstein model is of the order of 10$^{-4}$ and, in the absence of a quantitative theory, we cannot claim that PuCoGa$_5$ is at the verge of a critical valence transition on the basis of our results.

Our attempts to separate electronic and vibrational contributions to the observed thermal expansion failed. In PuCoGa$_5$ T$_c$ is relatively high, whereas its linear specific heat capacity is relatively small compared to other heavy-fermion materials. As a consequence, the phonon contribution to the thermal expansion cannot be considered as a small correction as in many heavy-fermion superconductors where the critical temperature is in the sub-kelvin range and the specific heat Sommerfeld coefficient $\gamma$ is very high. Therefore, in the absence of a precise estimate of the vibrational term, a reliable separation of the different contributions to the thermal expansion was not feasible.

\section{Conclusions}
X-ray diffraction measurements with a resolution of $\Delta d/d \sim$10$^{-6}$ in the lattice spacing show that the tetragonal symmetry exhibited by the PuCoGa$_{5}$ unconventional superconductor is preserved down to 2 K, well below the critical temperature T$_{c}$ = 18.7 K. The lattice thermal expansion is isotropic down to 150 K, and anisotropic for lower temperatures. This gives a $c/a$ ratio that decreases with increasing $T$ to become almost constant above $\sim$ 150 K. The volume thermal expansion coefficient $\alpha_{V}$ has a jump at $T_{c}$, a factor$\sim$ 20 larger than the change predicted by the Ehrenfest relation. At low temperatures, the expansion of the unit cell volume deviates from the curve corresponding to a simple one-phonon Gr\"{u}neisen-Einstein model and shows, below T$_{c}$, a continuous linear shrinking of the volume. In the case of the CeRu$_{2}$Si$_{2}$ Kondo system, a similar trend has been attributed to critical valence fluctuations. Although the deviations observed for PuCoGa$_{5}$ are about ten times larger than the statistical errors, in the absence of a quantitative theory it is not possible to establish the occurrence of critical valence fluctuations near T$_{c}$. The determination of thermal expansion along the main directions  in PuCoGa$_5$ single crystals with a technique affording higher sensitivity and a higher density of experimental points like dilatometry would be welcome to study more precisely this anomaly at T$_c$, confirm and refine ours observations and stimulate theoretical works.

\section{Acknowledgement}
We thank A. Fitch, J. M. Lawrence, and E. D. Bauer for stimulating discussions and advice. We are grateful to P. Colomp of the ESRF radioprotection services for his cooperation during the execution of the experiment.

\bibliography{PuCoGa5}

\end{document}